\documentclass[a4paper,fleqn]{cas-sc}
\usepackage{placeins}

\usepackage[numbers,sort&compress]{natbib}

\usepackage[utf8]{inputenc}
\usepackage[T1]{fontenc}

\usepackage{amsmath,amssymb,amsfonts}
\usepackage{textcomp}
\usepackage{bm}

\usepackage{graphicx}
\usepackage{subcaption}
\usepackage{booktabs}
\usepackage{multirow}
\usepackage{array}
\usepackage{multicol}
\usepackage{float}

\usepackage{algorithm}
\usepackage{algpseudocode}

\usepackage{url}
\usepackage{xcolor}

\begin{document}
\let\WriteBookmarks\relax
\def\floatpagepagefraction{1}
\def\textpagefraction{.001}

\shorttitle{Power-Aware Cognitive Radar Multi-target Tracking}
\shortauthors{I.~Bouhou et al.}

\title[mode=title]{Power-Aware Cognitive Radar Multi-target Tracking Under Unknown Disturbances}

\author[1,3]{Imad Bouhou}
\cormark[1]
\ead{imad.bouhou@outlook.com}
\credit{Conceptualization, Investigation, Methodology, Software, Validation, Writing -- original draft,  Writing - review and editing}

\author[2]{Stefano Fortunati}
\ead{stefano.fortunati@telecom-sudparis.eu}
\credit{Funding acquisition, Supervision, Writing -- review and editing}

\author[3]{Leila Gharsalli}
\ead{leila.gharsalli@ipsa.fr}
\credit{Funding acquisition, Supervision, Writing -- review and editing}

\author[1]{Alexandre Renaux}
\ead{alexandre.renaux@universite-paris-saclay.fr}
\credit{Project administration,  Supervision, Writing -- review and editing}

\affiliation[1]{organization={Universit\'e Paris-Saclay, CNRS, CentraleSup\'elec, Laboratoire des Signaux et Syst\`emes},
                addressline={3 rue Joliot-Curie},
                postcode={91190},
                city={Gif-sur-Yvette},
                country={France}}

\affiliation[2]{organization={SAMOVAR, T\'el\'ecom SudParis, Institut Polytechnique de Paris},
                postcode={91120},
                city={Évry},
                country={France}}

\affiliation[3]{organization={DR2I-IPSA},
                city={Ivry-sur-Seine},
                country={France}}

\cortext[1]{Corresponding author}

\begin{abstract}
This work presents a cognitive radar (CR) framework designed to track multiple aircraft under unknown disturbances using massive multiple-input multiple-output (MMIMO) systems. Since uniform power allocation is suboptimal across varying signal-to-noise ratios (SNRs), we couple an adaptive waveform design driven by Partially Observable Monte Carlo Planning (POMCP). By assigning an independent POMCP tree to each target, the system efficiently predicts target states. These predictions inform a constrained optimization problem that actively directs transmit energy toward weaker targets while maintaining sufficient power for stronger ones. Results confirm that the proposed POMCP method improves the detection probability for low-SNR targets from 0.6 to nearly 0.9, and yields more accurate tracking of the weakest target than a non-adaptive orthogonal waveform or a cognitive uniform-power POMCP baseline.
\end{abstract}


\maketitle

\section{Introduction}

Modern radar systems are indispensable components in a growing spectrum of applications, from autonomous driving and air traffic control to defense surveillance and intelligent sensing \cite{soni}. In these increasingly dense and dynamic environments, the ability to reliably detect and track multiple targets is paramount. Massive Multiple-Input Multiple-Output (MMIMO) radars have emerged as a key enabling technology, leveraging a vast number of antennas to provide unparalleled spatial resolution, enhance parameter estimation accuracy, and offer inherent robustness to interference \cite{massive_mimo_wald}. However, advanced hardware alone is insufficient. To operate effectively in complex scenarios, radar systems must evolve from rigid, pre-programmed transmission schemes toward an Artificial Intelligence driven paradigm in which the radar acts as an autonomous agent that adapts its transmissions online to its environment.

This need has led to the development of cognitive radar, defined by a closed-loop perception–action cycle \cite{cr_haykin_way_of_future}. A cognitive radar perceives the environment, learns from observations, and adapts its future transmissions (encompassing waveform, power allocation, and beam direction) to optimize performance. Its decision-making process can naturally be modeled as a Partially Observable Markov Decision Process (POMDP), a well-established framework in sequential decision-making under uncertainty. In the POMDP formulation, the radar acts as an agent that selects actions to maximize long-term reward, while maintaining a probabilistic belief over hidden target states inferred from noisy and incomplete measurements.

Many approaches have been developed to improve the radar decision-making engine. Classical probabilistic methods, such as Bayesian adaptive detection and tracking \cite{bell_cr}. Machine learning methods have been explored for tasks such as maritime target detection \cite{maritime} and, more recently, Deep Reinforcement Learning (DRL) has been applied to airborne radar trajectory optimization  \cite{drl_cr} and air-combat decision-making under uncertainty \cite{aircraft}. Deep learning has been used for joint cognitive radar-communication waveform design \cite{munir}. While powerful, DRL methods often function as black boxes, require extensive offline training on large representative datasets, and can struggle to generalize to unknown environmental statistics, which limits their applicability in real-time radar systems. To address these limitations, recent literature has explored online planners. Notably, our previous work successfully deployed the POMCP algorithm \cite{pomcp} for single-target tracking in MMIMO radar \cite{mimo_pomcp}. As a Monte Carlo tree search (MCTS)-based online reinforcement learning algorithm, POMCP constructs action policies through simulated rollouts from the current belief state, enabling real-time adaptation without pre-collected data. While the POMCP framework has proven highly effective and interpretable for isolated targets, extending this cognitive architecture to multi-target environments raises new challenges. A naive joint formulation would require the radar to reason over all targets simultaneously, which becomes increasingly difficult as the number of targets grows. To address this, this work extends the online POMCP framework into a decentralized cognitive radar architecture for multi-target detection and tracking. Rather than searching over a joint multi-target action space, our proposed framework assigns an independent planner to each target and combines this decentralized planning with an adaptive power allocation strategy. 

The key contributions of this work are threefold:
\begin{enumerate}
    \item We extend the online POMCP cognitive radar framework of \cite{mimo_pomcp} from single-target to multi-target settings under the standard non-intersection assumption, which yields an exact decomposition into $M$ independent POMCP trees and avoids the need to search over a joint multi-target action space.
    \item We integrate an adaptive power-allocation layer in which the per-step transmit-waveform design is cast as a constrained optimization driven by particle-filter-based predictions of target ranges and expected received powers.
    \item We empirically demonstrate that the proposed allocation preferentially improves detection and tracking of the weakest, low-SNR target without degrading the performance of the high-SNR targets, in a setting with completely unknown disturbance statistics.
\end{enumerate}

We validate the proposed framework through simulations featuring multiple targets with diverse Signal-to-Noise Ratio (SNR) profiles. While DRL has shown promise in adaptive tracking \cite{drl_cr}, such methods fundamentally require extensive offline training on representative datasets. In environments characterized by an unknown disturbance distribution $p_C$, generating a valid training set is physically impractical. Our framework instead utilizes online POMCP to construct action policies directly from the current belief state in real-time, completely bypassing the need for prior environmental knowledge. Robustness to the unknown disturbance distribution $p_C$ was empirically established for the single-target setting in \cite{mimo_pomcp} across several disturbance models, and it is inherited here since the detection statistic and the belief-update mechanism remain unchanged. Consequently, this work provides an online, training-free baseline for adaptive power allocation in settings where obtaining representative offline training data is difficult. To demonstrate the survival and precision of this architecture in unknown clutter, we benchmark our power-aware framework against classical orthogonal and uniform energy transmission strategies.

\section{Problem Formulation}
\label{sec:problem_formulation}
This section provides a brief overview of the system model, which is identical to that presented in \cite{mimo_pomcp}. We consider a MMIMO radar, equipped with a large number of antennas.

\subsection{System Model}
The Massive MIMO radar is equipped with $N_T$ transmit and $N_R$ receive physical antennas, resulting in $N = N_T N_R$ virtual spatial antenna channels. The radar's field of view is divided into $L_\theta$ angle bins. At each time step $t$, the system scans the environment by transmitting a waveform. The detection problem for a specific angle bin $l$ at time $t+1$ is formulated under two hypotheses:
\begin{equation}
    \begin{split}
        H_0 &: \mathbf{y}_{t+1,l} = \mathbf{c}_{t+1,l},\\
        H_1 &: \mathbf{y}_{t+1,l} = \alpha_{t+1,l}\mathbf{v}_{t,l} + \mathbf{c}_{t+1,l}.
    \end{split}
    \label{ht}
\end{equation}

In this multi-target scenario, it is possible that multiple angle bins will correspond to the $H_1$ hypothesis. Here, $\mathbf{c}_{t+1,l} \in \mathbb{C}^{N}$ is a random vector representing the disturbance, possessing an unknown probability density function $p_C$. Its auto-correlation function is assumed to exist and decay at least at a polynomial rate, as noted in \cite{massive_mimo_wald}. The term $\alpha_{t+1,l} \in \mathbb{C}$ is an unknown deterministic scalar that accounts for the Radar Cross-Section (RCS) and two-way path loss. The vector $\mathbf{v}_{t,l}$ is defined as:
\begin{equation}
    \mathbf{v}_{t,l} = (\mathbf{W}_t^T \mathbf{a}_T(\theta_l))\otimes\mathbf{a}_R(\theta_l) \in \mathbb{C}^{N},
\end{equation}
where $\mathbf{a}_R(\theta_l)$ and $\mathbf{a}_T(\theta_l)$ are known receive and transmit steering vectors, respectively. The waveform matrix $\mathbf{W}_t \in \mathbb{C}^{N_T \times N_T}$ is selected to distribute the transmit energy across the chosen set of angle bins $\Theta$, while adhering to a total transmit power constraint $P_T$.
To handle the hypothesis testing problem in \eqref{ht}, we adopt the robust Wald-type test introduced in \cite{massive_mimo_wald} as:
\begin{equation}
    \Lambda_{t+1,l} = 2 |\hat{\alpha}_{t+1,l}|^2 \frac{||\mathbf{v}_{t,l}||^4}{\mathbf{v}_{t,l}^H \widehat{\mathbf{\Sigma}}_{t+1,l} \mathbf{v}_{t,l} } \underset{H_0}{\overset{H_1}{\gtrless}} \lambda,
\end{equation}
where $\widehat{\mathbf{\Sigma}}_{t+1,l}$ is the estimate of the disturbance covariance given in \cite[eq. (23)]{massive_mimo_wald}, and $\hat{\alpha}_{t+1,l} = (\mathbf{v}_{t,l}^H \mathbf{y}_{t+1,l})/||\mathbf{v}_{t,l}||^2$ is an estimate of $\alpha_{t+1,l}$. The closed-form expressions for the probability of detection and false alarm can be found in \cite{massive_mimo_wald}.

\section{Proposed Cognitive Architecture}
This section outlines the modifications made to the cognitive radar's design to handle multiple targets. Let $M > 1$ denote the number of targets in the environment. A brief reminder on the POMDP and the POMCP can be found in \cite{mimo_pomcp}.

A naive joint formulation for $M$ targets would require selecting a combination of angle bins for all targets simultaneously, which becomes increasingly hard as the number of targets grows. However, under the standard non-intersection assumption, the multi-target tracking problem decomposes exactly into $M$ independent single-target POMDPs. This decomposition is a structural consequence of the non-intersection assumption rather than an algorithmic contribution in itself; we exploit it by assigning an independent POMCP tree to each target, so that each target is tracked by a separate single-target search rather than through a single joint search. The substantive contribution of the present work is the adaptive power-allocation layer built on top of this decomposition.

\subsection{State Space}
The state space for multiple targets consists of the combined positions and velocities of all targets. At time step $t$, the state of the $m$-th target is defined as:
\begin{equation}
    \mathbf{s}_{t}^{(m)} = [x_t^{(m)}, V_{x,t}^{(m)}, y_t^{(m)}, V_{y,t}^{(m)}]^T,
\end{equation}
where $[x_t^{(m)}, y_t^{(m)}]$ and $[V_{x,t}^{(m)}, V_{y,t}^{(m)}]$ are the position and velocity vectors of the $m$-th target, respectively.

The dynamics of each target are described by $\mathbf{s}_{t+1}^{(m)} = \mathbf{A}\mathbf{s}_t^{(m)} + \mathbf{G}\mathbf{w}_t^{(m)}$, where the state transition matrix is $\mathbf{A} = \text{blockdiag}(\mathbf{A}_b, \mathbf{A}_b)$ with $\mathbf{A}_b = \left[\begin{smallmatrix} 1 & \Delta t \\ 0 & 1 \end{smallmatrix}\right]$. The noise matrix is $\mathbf{G} = \text{blockdiag}(\mathbf{G}_b, \mathbf{G}_b)$ with $\mathbf{G}_b = \left[\begin{smallmatrix} \Delta t^2 / 2 \\ \Delta t \end{smallmatrix}\right]$. The process noise $\mathbf{w}_t^{(m)}$ is i.i.d. Gaussian, $\mathbf{w}_t^{(m)} \sim \mathcal{N}(\mathbf{0}_2, \sigma_s^2 \mathbf{I}_2)$, where $\sigma_s$ is the standard deviation.

\subsection{The unweighted particle filter}
\label{subsec:pfpf}
The particle filter in this work serves two main roles. First, it updates the belief set at each iteration as new observations arrive, as explained in \cite{pomcp}. Second, it predicts the target’s future range. The first role ensures that POMCP continues to function and converge, while the second supports the radar’s power allocation strategy, guaranteeing that each target receives an appropriate amount of power, neither excessive nor insufficient.

At time step $t$, the radar has a history of observations and actions $h_t$ and can build an approximation of the posterior $b(.|h_t)$, which is defined by the set $B_t$. The prediction step using the particle filter translates to computing $\mathbb{E}( \mathbf{s}_{t+1}|h_t)$, which is defined as follows \cite{mimo_pomcp}:
\begin{align}
    \mathbb{E}(\mathbf{s}_{t+1}|h_t)
    &\approx \frac{1}{|B_t|} \sum_{\mathbf{s} \in B_t} \mathbb{E}(\mathbf{s}_{t+1}|\mathbf{s}_t=\mathbf{s}).
    \label{eq:pf-pred}
\end{align}

\subsection{Action Space}
\label{action_space}
The works of \cite{rl_mimo_aya} use a uniform power allocation waveform, denoted $\mathbf{W}_{\text{uni}}$, which assigns equal energy to all angle bins regardless of target RCS. In contrast, the radar here not only selects multiple angle bins but also distributes transmit energy optimally across targets by estimating their RCS values using the equation \ref{eq:pf-pred}.

At each time step $t$, the radar chooses an action $a_t$, defined as a set of angle bins $(\theta_t^{(m)})_{m=1}^{M}$, where $\theta_t^{(m)}$ is the bin assigned to target $m$, selected from $L_\theta$ possible bins. Using these bins and estimated target powers, the radar computes the waveform matrix to distribute its transmit energy optimally. For multiple targets, the optimal waveform design accounts for both target angles and estimated powers, unlike the uniform transmission in \cite{rl_mimo_aya}. Following \cite{li_wang}, the radar maximizes the minimum weighted beam pattern across all targets:
\begin{align}
    \max_{\mathbf{R}} \quad & \min_{m \in \{1,\ldots,M\}} \ \delta_t^{(m)} \, \mathbf{a}_T^T(\theta_t^{(m)}) \, \mathbf{R} \, \mathbf{a}_T^*(\theta_t^{(m)}) \notag \\
    \text{subject to} \quad 
    & \text{Tr}(\mathbf{W}\mathbf{W}^H) = P_T, \notag \\
    & \mathbf{R} = \mathbf{W}\mathbf{W}^H.
    \label{w_new_multiple}
\end{align}

The parameter $\delta_t^{(m)}$ denotes the expected power of target $m$ at the next step. Let $\tilde{R}_{t+1,m}$ be its expected range; from the radar equation, $\delta_t^{(m)} = 1/\tilde{R}_{t+1,m}^{4}$.  
At time $t$, the radar maintains a belief set $B_t^{(m)}$ for each target and predicts the next state using the unweighted particle filter of section \ref{subsec:pfpf}. The resulting waveform $\mathbf{W}_{\delta}$ from \eqref{w_new_multiple} adapts the transmitted energy to the estimated target strengths, thereby improving detection and tracking in multi-target settings.  

The equation \eqref{eq:pf-pred} is useful to compute the coefficients $(\delta_t^{(m)})_{m =1}^M$ and consequently solve the optimization problem \eqref{w_new_multiple} using CVX toolbox \cite{cvx}, after the angle bin selection by the POMCP.

\subsection{Observation Space}
At time step $t$, the radar performs an action $a_t$ corresponding to a set of angle bins $a_t = \{ \theta_t^{(1)}, \theta_t^{(2)}, \ldots, \theta_t^{(M)}\}$ and estimates target power coefficients $\{ \delta_t^{(1)}, \delta_t^{(2)}, \ldots, \delta_t^{(M)}\}$. These parameters are used to compute the waveform vector $\mathbf{v}_{t,l}$ using the waveform obtained as the solution of the constrained optimization problem \eqref{w_new_multiple}. The radar then receives an observation, which is either the estimated parameter $|\alpha_{t+1, l}^{(m)}|$ for each detected target, or an empty observation otherwise:
\begin{equation}
    o_{t+1}^{(m)} =
    \begin{cases}
        |\hat{\alpha}_{t+1,l}^{(m)}| & \text{if } \Lambda_{t+1,l}^{(m)} \geq \lambda, \\
        \emptyset & \text{otherwise},
    \end{cases}
    \label{alpha_obs}
\end{equation}
where $\Lambda_{t+1,l}^{(m)}$ is the detection test statistic for the $m$-th target and $\lambda$ is the detection threshold.

Consistent with the radar equation, the true parameter $|\alpha_{t+1,l}^{(m)}|$ is inversely proportional to the square of the range $R_{t+1,m}$ between the target and the radar:
\begin{equation}
    |\alpha_{t+1,l}^{(m)}| \propto 1/R_{t+1,m}^2.
    \label{eq:alpha_range_relation}
\end{equation}

As shown in \cite{massive_mimo_wald}, the estimated parameter $\hat{\alpha}_{t+1,l}^{(m)}$ is asymptotically distributed as a complex Gaussian:
\begin{equation}
    (\hat{\alpha}_{t+1,l}^{(m)} - \alpha_{t+1,l}^{(m)})/\widehat{\sigma}_{t,l} \underset{N \to \infty}{\sim} \mathcal{CN}(0,1),
    \label{alpha_asymptotic}
\end{equation}
where 
\begin{equation}
    \widehat{\sigma}_{t,l} = \sqrt{\mathbf{v}_{t,l}^H \widehat{\mathbf{\Sigma}}_{t+1,l}\mathbf{v}_{t,l}}/\|\mathbf{v}_{t,l}\|^2.
    \label{standard_deviation}
\end{equation}

Because the POMCP algorithm requires discrete observations, the continuous observations must be mapped into a discrete space using a step size $\beta_l$. To determine a statistically sound value, we rely on the fact that the squared estimation error $|\hat{\alpha}_{t+1,l} - \alpha_{t+1,l}|^2$ asymptotically follows an exponential distribution characterized by the parameter $\widehat{\sigma}_{t,l}^2$. By evaluating the cumulative distribution function of this exponential distribution, we find that setting the step size to $\beta_{l} = \sqrt{3}\widehat{\sigma}_{t,l}$ guarantees the true parameter falls within the discretization bin with a probability of $0.95$. This approach, similar to the one in \cite{mimo_pomcp}, ensures that the POMCP tree search operates on highly reliable discrete bounds without needing the exact probability density function. In a single-target scenario with uniform power allocation, the system only generates $L_{\theta}$ possible waveforms, allowing for the offline pre-computation of all standard deviations in \eqref{alpha_asymptotic}. However, in our multi-target framework, the adaptive waveform matrix $\mathbf{W}_t$ is generated dynamically by solving the optimization problem in \eqref{w_new_multiple}. Because this optimization is parameterized by the continuous target power coefficients $\delta_t^{(m)}$, the radar can generate an infinite set of possible waveforms. Consequently, $\mathbf{v}_{t,l}$ and the resulting $\widehat{\sigma}_{t,l}$ cannot be precomputed in a finite lookup table. To resolve this, the standard deviations are instead updated online after each new detection, as detailed in Section~\ref{simulation_model}.

\subsection{Reward Function }
The reward function is designed to incentivize the radar to accurately detect and track targets within the environment. In the POMDP framework, the reward function depends on the chosen action $a_t^{(m)}$, and the subsequent state $\mathbf{s}_{t+1}$. The action $a_t^{(m)}$ is the choice of $\theta_t^{(m)}$ for each target $m$, and $\theta_{\mathbf{s}_{t+1}}^{(m)}$ denotes the true angle bin of the $m$-th target at time $t+1$. To encourage precise prediction of the target's future angle, the reward function is defined the same way as in  \cite{mimo_pomcp}:
\begin{equation}
    r_{t} = \mathbf{1}\{\theta_t^{(m)} = \theta_{\mathbf{s}_{t+1}}^{(m)}\}. 
    \label{reward_function}
\end{equation}

\subsection{Simulation Model}
\label{simulation_model}

\begin{algorithm}[H]
\caption{Generator $\mathcal{G}(\mathbf{s}_t, a_t)$.}
\label{generator}
\begin{algorithmic}[1]
\Require $\mathbf{s}_t=(x_t, V_{x,t}, y_t, V_{y,t})^T$, action $a_t$ and $\widehat{\sigma}$.
\State $\mathbf{s}_{t+1} \gets \mathbf{A}\mathbf{s}_t + \mathbf{G}\mathbf{w}_t$
\State $\theta_{\mathbf{s}_{t+1}}\gets \texttt{GetAngleBin}(\mathbf{s}_{t+1})$
\State $l_{t} \gets \texttt{GetAngleBin}(a_t)$
\State $\alpha_{t+1} \gets \texttt{GetRCS}(\mathbf{s}_{t+1})$ 
\State $\widehat{\alpha}_{t+1} \gets \mathcal{CN}(\alpha_{t+1}, \widehat{\sigma}^2)$ ; $\Lambda_{t+1} \gets \frac{2 |\widehat{\alpha}_{t+1}|^2}{\widehat{\sigma}^2}$
\If{$l_{t} \neq \theta_{\mathbf{s}_{t+1}}$} $o_{t+1} \gets \emptyset$
\ElsIf{$l_{t} = \theta_{\mathbf{s}_{t+1}}$}
    \If{$\Lambda_{t+1} \geq \lambda$} $o_{t+1} \gets |\widehat{\alpha}_{t+1}|$
    \Else \ $o_{t+1} \gets \emptyset$
    \EndIf
\EndIf
\State $r_{t} \gets \mathbf{1}\{l_t = \theta_{\mathbf{s}_{t+1}}\} $
\State \Return $(\mathbf{s}_{t+1}, o_{t+1}, r_{t})$
\end{algorithmic}
\end{algorithm}

POMCP relies on a black-box generator $\mathcal{G}(\mathbf{s},a)=(\mathbf{s}',o,r)$. In the single-target case, the $L_\theta$ actions allow pre-computation of $(\widehat{\sigma}_l)_{l=1}^{L_\theta}$. In the multi-target case, the action space and continuous nature of the parameters $\delta_t^{(m)}$ make such pre-computation intractable. We therefore update $\widehat{\sigma}$ online only after detection. When $\Lambda_{t+1,l}^{(m)}>\lambda$, the corresponding $\widehat{\sigma}^{(m)}$ is computed and stored. This approximation is motivated by the continuity of the disturbance's Power Spectral Density (PSD), which suggests that the corresponding $\widehat{\sigma}^{(m)}$ values should vary smoothly across nearby angle bins.

As established, each target $m$ uses an independent POMCP tree with generator $\mathcal{G}^{(m)}$, which maintains and updates its own $\widehat{\sigma}^{(m)}$ from the most recent detection. Algorithm~\ref{generator} summarizes $\mathcal{G}^{(m)}(\mathbf{s}_t^{(m)},a_t^{(m)})$. The \texttt{GetAngleBin} function returns the angle bin, \texttt{GetRCS} computes $\alpha_{t+1}^{(m)}=|\alpha_{t+1}^{(m)}|e^{j\phi}$ with $\phi\sim\mathcal{U}(0,2\pi)$, and $o_{t+1}^{(m)}=\emptyset$ when the selected bin does not match the true future angle $\theta_{t+1}^{(m)}$.

\subsection{Online Learning-Based Radar Design}
The cognitive radar initially transmits an orthogonal waveform matrix, $\mathbf{W}_{\text{ort}} = \sqrt{\frac{P_T}{N_T}}\mathbf{I}_{N_T{}}$, as detailed in \cite{mimo_pomcp}, until all targets are detected. Upon detection, target coordinates are estimated from observations, and velocities are uniformly initialized within $[-V_{\text{max}}, V_{\text{max}}]$ where $V_{\text{max}}$ is some predefined maximum velocity value. During this initial phase, the standard deviation associated with the detection angle bin, essential for the asymptotic relation in \eqref{alpha_asymptotic}, is computed and stored for each target.

The full cognitive radar design for multiple targets, including the use of POMCP, is presented in Algorithm \ref{cr_design}.

\begin{algorithm}
\caption{Cognitive radar design for multiple targets.}
\label{cr_design}
\begin{algorithmic}[1]
\Require $N_{\text{sim}}$ 
\Require $\{B_{0}^{(m)}\}_{m=1}^M$ \Comment{Initial belief sets for the targets.}
\Require $\{\mathcal{G}^{(m)}\}_{m=1}^M$ 
\Require $\{\widehat{\sigma}^{(m)}\}_{m=1}^M$ 
\Require $\beta^{(m)} = \sqrt{3} \widehat{\sigma}^{(m)} \text{ for } m=1, \dots, M$ 
\For{each time step $t=0,..,T_{\text{max}}-1$}
    \For{each detected target $m=1, \dots, M$}
        \State $a_t^{(m)} \gets \texttt{POMCP.Solve}(N_{\text{sim}}, B_t^{(m)})$. 
    \EndFor
    \State Compute the targets' power $\{\delta_t^{(m)}\}_{m=1}^M$ using \eqref{eq:pf-pred}.
    \State Compute the waveform matrix $\mathbf{W}_t$ based on $\{a_t^{(m)}\}_{m=1}^M$ and $\{\delta_t^{(m)}\}_{m=1}^M$ by solving \eqref{w_new_multiple}.

    \State Receive the signals $\{\mathbf{y}_{t+1,l}^{(m)}\}_{m=1}^M$ and observe $\{o_{t+1}^{(m)}\}_{m=1}^M$ from \eqref{alpha_obs}.
    \For{each detected target $\Lambda_{t+1,l}^{(m)} > \lambda$}
        \State Update $\widehat{\sigma}^{(m)}$ for $m$-th target's generator with the newly observed standard deviation and  $\beta^{(m)} = \sqrt{3} \widehat{\sigma}^{(m)}$.
    \EndFor
    \For{all $m=1,\cdots,M$}
    \State $B_{t+1}^{(m)} \gets \texttt{UpdateBelief}(B_t^{(m)}, a_t^{(m)}, o_{t+1}^{(m)})$.
    \EndFor
\EndFor
\end{algorithmic}
\end{algorithm}

\section{Simulations}
In our simulation setup, we use the same parameters as the ones used in \cite{mimo_pomcp}. Regarding the targets, two primary assumptions are made. First, the radar is presumed to have prior knowledge of the total number $M$ of targets it needs to detect. Consequently, the radar will select $M$ distinct angle bins during each iteration. Second, we assume that targets never spatially overlap; that is, they do not share the same angle bin or intersect with each other. 
This assumption is standard in multi-target tracking literature. Handling potential target merging and splitting events falls under the scope of data association methods \cite{jpda}, which are orthogonal to the contributions of this work and constitute a natural direction for future extension. We further note that, within our framework, the unweighted particle filter is used both for belief propagation and for predicting target range; alternative tracking layers were evaluated in \cite{mimo_pomcp} and did not improve performance over the present configuration, which is why the comparison in this work is restricted to waveform and power-allocation strategies sharing the same underlying tracker. The objective of this simulation is twofold: first, to evaluate the proposed algorithm in a simplified multi-target embedded in an unknown disturbance; and second, to compare uniform energy transmission with transmission guided by predicted target power, with particular attention to the weakest target trajectory.

In this simulation, three targets are considered, with their initial states defined as follows:
\begin{align*}
\mathbf{s}_0^{(1)} &= [20 \text{km}, 0.05 \text{km/s}, -60 \text{km}, 0.01 \text{km/s}]^T, \\
\mathbf{s}_0^{(2)} &= [60 \text{km}, 0.20 \text{km/s}, 7.5 \text{km}, 0.10 \text{km/s}]^T,\\
\mathbf{s}_0^{(3)} &= [5 \text{km}, 0.05 \text{km/s}, 60 \text{km}, 0.01 \text{km/s}]^T.
\end{align*}
The standard deviation of the noise processes is $\sigma_s = 0.004 \text{km/s}^2$. On average, Target 1 has a SNR trajectory that begins at \(-12\,\mathrm{dB}\) and decreases to \(-13\,\mathrm{dB}\), Target 2 starts at \(-12\,\mathrm{dB}\) and drops to \(-24\,\mathrm{dB}\), while Target 3 starts at  \(-12\,\mathrm{dB}\) and drops to \(-14.5\,\mathrm{dB}\). The objective is to evaluate whether the algorithm can better detect Target 2 when using \(\mathbf{W}_{\delta}\) compared to using the uniform transmission strategy using \(\mathbf{W}_{\text{uni}}\).

The radar has the following configuration: number of virtual spatial channels $N = N_T N_R = 10^4$, number of angle bins $L_{\theta} = N_T = 100$, total transmit power $P_T = 1$, and false alarm probability $P_{FA} = 10^{-4}$.
To optimize experimental runtime, we configured the search trees with $12\ 000$ particles ($N_p$) and $12\ 000$ simulations ($N_{\text{sim}}$). The exploration-exploitation parameter, $c$, was set to $\sqrt{2}$. Results are averaged over $100$ Monte Carlo runs.

\begin{figure}[!htbp]
    \centering
    \includegraphics[width=\linewidth]{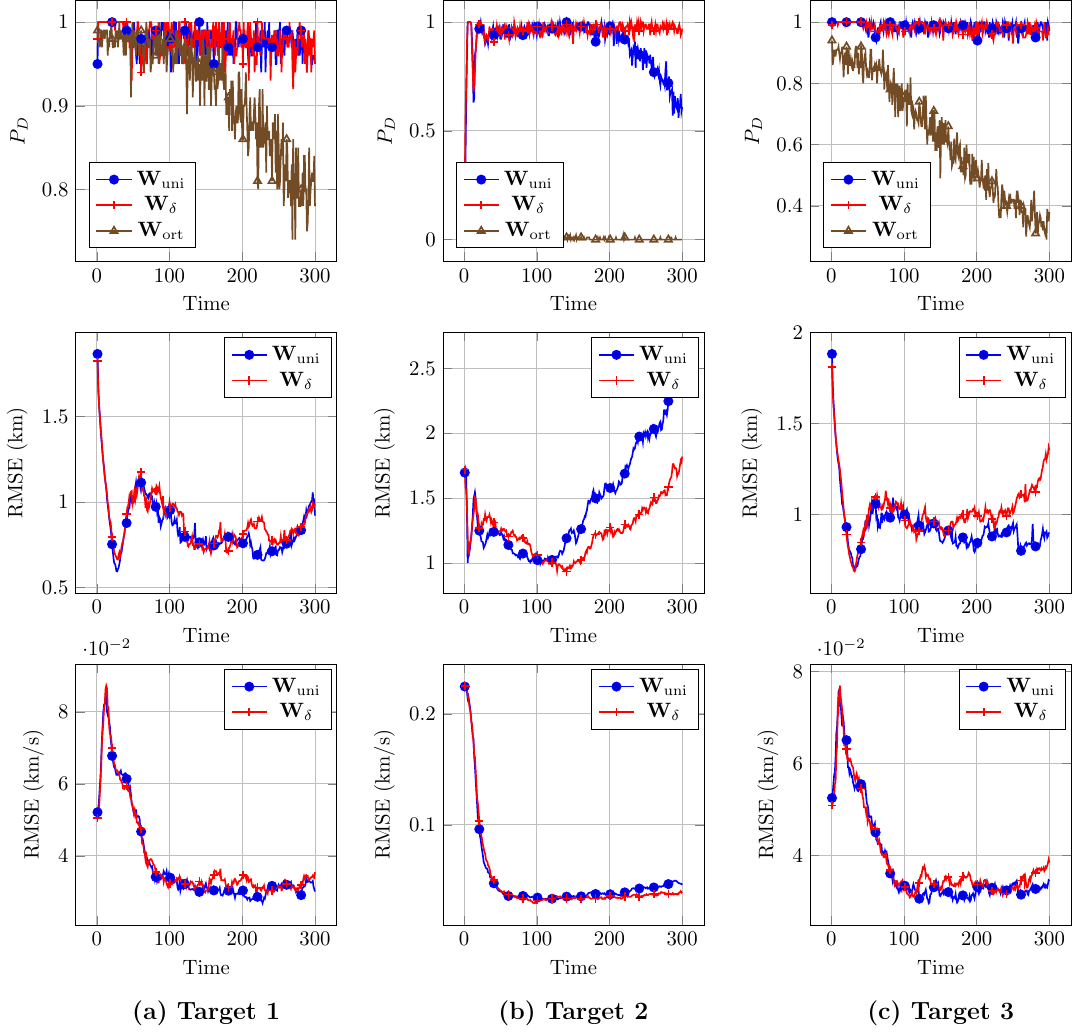}
    \caption{Performance metrics for all tracked targets.}
    \label{fig:all_targets}
\end{figure}

The simulation results are presented in Figure~\ref{fig:all_targets}. The main observation is that the benefit of the proposed power-aware allocation is primarily concentrated on Target~2, which corresponds to the most challenging low-SNR trajectory. For this target, $\mathbf{W}_{\delta}$ sustains a high detection probability over the full horizon, whereas $\mathbf{W}_{\mathrm{uni}}$ deteriorates noticeably at later times and $\mathbf{W}_{\mathrm{ort}}$ rapidly collapses. Concretely, at the final time steps the detection probability of Target~2 drops to approximately $0.6$ under $\mathbf{W}_{\mathrm{uni}}$, while the proposed $\mathbf{W}_{\delta}$ maintains it near $0.9$ (Fig.~\ref{fig:all_targets}), an improvement of roughly $50\%$ for the weakest target. By contrast, for Targets~1 and~3, both adaptive strategies, $\mathbf{W}_{\mathrm{uni}}$ and $\mathbf{W}_{\delta}$, achieve similarly high detection probabilities, while $\mathbf{W}_{\mathrm{ort}}$ degrades progressively with time. 
The tracking results are consistent with the detection trends. For Target~2, the proposed waveform yields a lower position RMSE than $\mathbf{W}_{\mathrm{uni}}$, particularly in the second half of the trajectory. This indicates that focusing on the weakest target improves tracking accuracy when detection becomes harder. 

However, for Targets~1 and~3, the difference in position RMSE between the two methods is small, and neither method is consistently better. Similarly, the velocity RMSE curves are very close for all three targets. Overall, the main benefit of the proposed method is keeping track of the weakest target, rather than improving the results for all targets and metrics.
This is consistent with the goal of the proposed design, which shares a limited amount of transmit power among multiple targets. Since the total power is fixed, allocating more energy to the weakest low-SNR target will not improve the results for all targets at once. Ultimately, this method demonstrates a highly favorable trade-off: it provides a substantial tracking benefit for the weakest target without causing disproportionate degradation to the detection of the others.

\subsection{Adaptive Waveform and Target 2 Tracking Behavior}

\begin{figure}[H]
    \centering
    \includegraphics[width=0.85\linewidth]{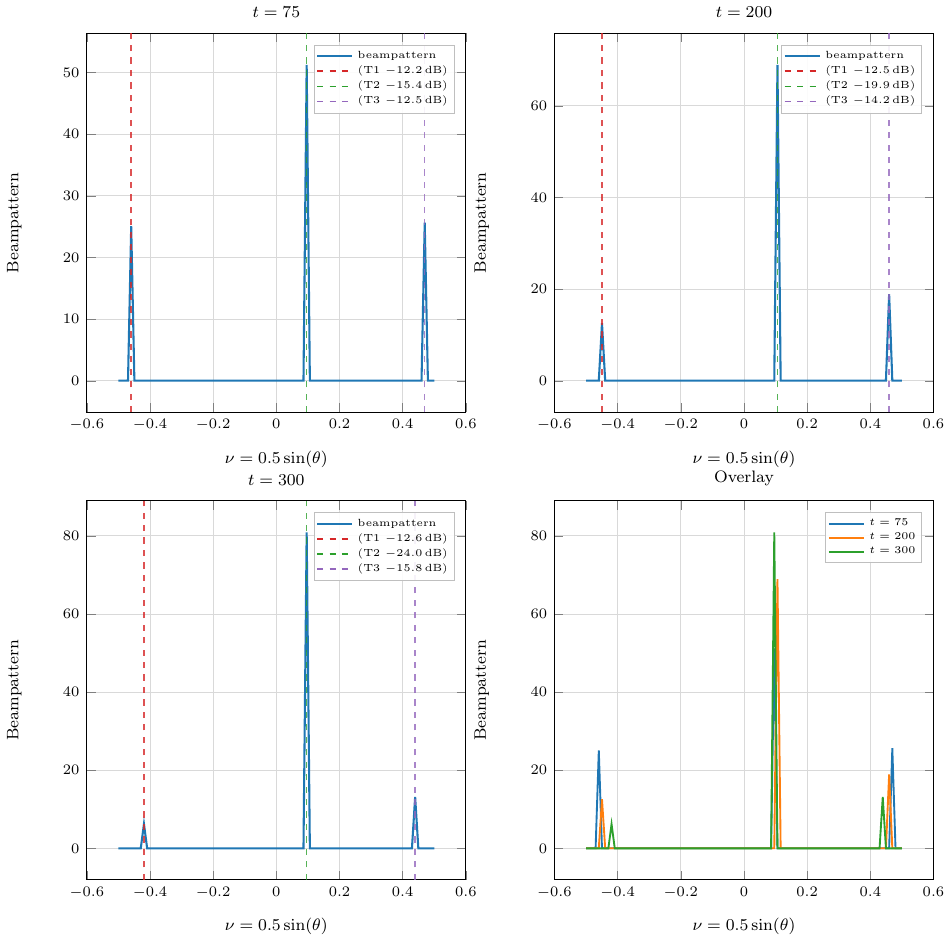}
    \caption{Adaptive beampatterns at snapshot times $t=75$, $t=200$, and $t=300$. The vertical lines indicate the true angle bins of the three targets along with their instantaneous SNRs. The radar dynamically focuses maximum gain on the weakest target (Target 2).}
    \label{fig:beampatterns}
\end{figure}

To better understand why the radar maintains this level of performance, we can examine how it adaptively distributes its spatial energy. Figure~\ref{fig:beampatterns} shows the transmitted beampatterns at different snapshot times. At an early stage of the simulation ($t=75$), Target 2 has a moderate SNR of -15.4 dB. As the target moves farther away, the signal quality decreases significantly, reaching -19.9 dB at $t=200$ and finally -24.0 dB. The figure highlights the response of the cognitive radar system: the optimization problem in \eqref{w_new_multiple} automatically generates a waveform with a sharper and stronger main lobe directed toward the predicted angle of Target 2. This adaptive focusing compensates for the increasing path loss. At the same time, the radar continues to allocate enough transmit energy to Targets 1 and 3, which maintain more favorable SNR values (-12.6 dB and -15.8 dB), ensuring reliable detection for all targets.

\begin{figure}[H]
    \centering
    \includegraphics[width=0.85\linewidth]{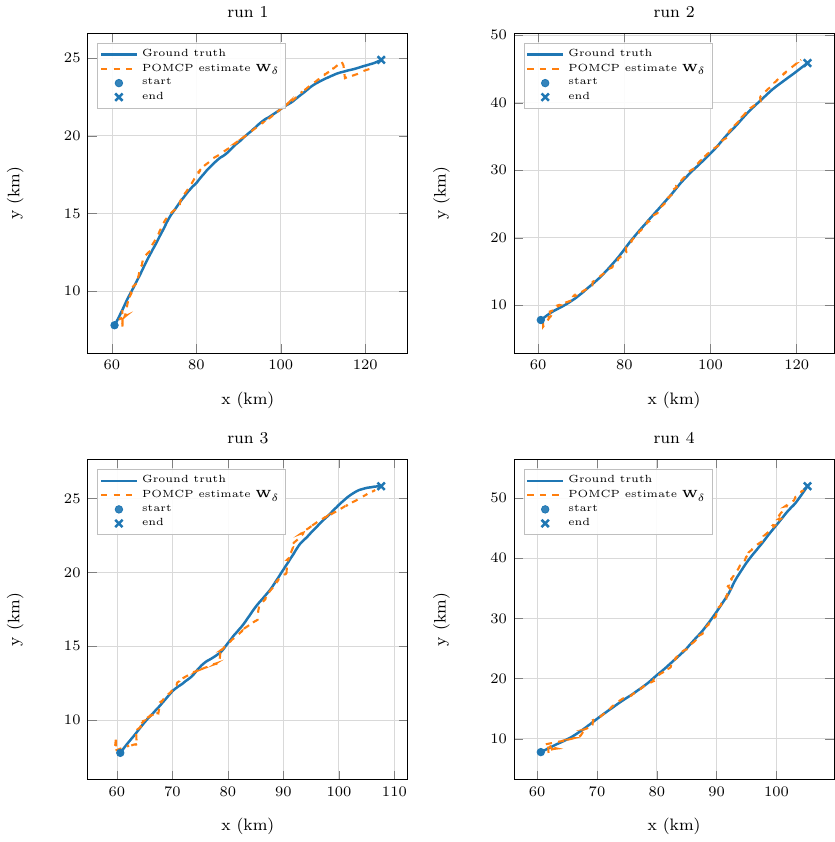}
    \caption{Target 2 trajectories across four independent Monte Carlo runs. Despite severe SNR degradation, the POMCP tracking estimate remains tightly coupled to the ground truth.}
    \label{fig:traj_target2}
\end{figure}

The tangible result of this adaptive allocation is shown in Figure~\ref{fig:traj_target2}, which isolates the trajectory of Target 2 across four independent Monte Carlo runs. Despite the target reaching an exceptionally low SNR of -24.0 dB, the POMCP estimate consistently adheres to the ground truth trajectory from the start point to the end point. This confirms that the dynamic waveform design successfully prevents track loss for the weakest target without requiring any prior knowledge of the disturbance distribution.

\FloatBarrier
\section{Conclusion}
This work extends the POMCP-based algorithm proposed in \cite{mimo_pomcp} to multi-target detection and tracking in massive MIMO radar, with dynamic power allocation based on target SNRs inspired by \cite{li_wang}. Simulations show improved low-SNR target detection and tracking over a cognitive uniform-power POMCP baseline. The current framework assumes a known number of spatially separated targets and is validated in a synthetic environment only, as no publicly available real-measurement dataset exists for massive MIMO radar. Future work will address the case of an unknown number of targets and targets with intersecting trajectories. Specifically, we aim to integrate advanced data association techniques, such as JPDA \cite{jpda}, directly into the POMCP framework to seamlessly resolve spatially overlapping targets and crossing trajectories.


\section*{Funding}
This work was supported by Institut Polytechnique des Sciences Avancées-Ivry-sur-Seine, France.


\section*{Data Availability}
No empirical data was used in this research. The data presented in this study were generated through Python-based software simulations. The simulation code used to reproduce these findings is currently available from the corresponding author upon request, and will be made publicly available on GitHub upon acceptance of the article.



\printcredits

\bibliographystyle{elsarticle-num}
\bibliography{bibt}

\end{document}